\documentclass[prd,showpacs, twocolumn,amsmath,amssymb]{revtex4}
\usepackage{amssymb}
\usepackage{graphicx}% Include figure files
\usepackage{dcolumn}% Align table columns on decimal point
\usepackage{bm}% bold math
\usepackage{amsfonts}
\usepackage{keyval,graphicx}
\usepackage{textcomp,wasysym}

\begin{document}

%%%%%%%%%%%%%%%%%%%%%%%%%%%%%%%%%%%%%%%%%%%%%%%%%%%%%%%%%%%%%%%%%%%%
\def\bs{$\backslash$} \def\bslp{\baselineskip}
\def\spg{\setcounter{page}} \def\seq{\setcounter{equation}}
\def\bd{\begin{document}} \def\ed{\end{document}}
\def\bmp{\begin{minipage}} \def\emp{\end{minipage}}
\def\bcc{\begin{center}} \def\ecc{\end{center}}     \def\npg{\newpage}
\def\beq{\begin{equation}} \def\eeq{\end{equation}} \def\hph{\hphantom}
\def\be{\begin{equation}} \def\ee{\end{equation}} \def\r#1{$^{[#1]}$}
\def\n{\noindent} \def\ni{\noindent} \def\pa{\parindent}
\def\hs{\hskip} \def\vs{\vskip} \def\hf{\hfill} \def\ej{\vfill\eject}
\def\cl{\centerline} \def\ob{\obeylines}  \def\ls{\leftskip}
\def\underbar#1{$\setbox0=\hbox{#1} \dp0=1.5pt \mathsurround=0pt
   \underline{\box0}$}   \def\ub{\underbar}    \def\ul{\underline}
\def\f{\left} \def\g{\right} \def\e{{\rm e}} \def\o{\over} \def\d{{\rm d}}
\def\vf{\varphi} \def\pl{\partial} \def\cov{{\rm cov}} \def\ch{{\rm ch}}
\def\la{\langle} \def\ra{\rangle} \def\EE{e$^+$e$^-$} \def\pt{p_{\rm t}}
\def\bitz{\begin{itemize}} \def\eitz{\end{itemize}}
\def\btbl{\begin{tabular}} \def\etbl{\end{tabular}}
\def\btbb{\begin{tabbing}} \def\etbb{\end{tabbing}}
\def\beqar{\begin{eqnarray}} \def\eeqar{\end{eqnarray}}
\def\\{\hfill\break} \def\dit{\item{-}} \def\i{\item}
\def\bbb{\begin{thebibliography}{9} \itemsep=-1mm}
\def\ebb{\end{thebibliography}} \def\bb{\bibitem}
\def\bpic{\begin{picture}(260,240)} \def\epic{\end{picture}}
\def\akgt{\cl{\bf ACKNOWLEDGMENTS}}
\def\fgn{\noindent{\bf\large\bf figure captions}}
%%%%%%%%%%%%%%%%  \begin{eqnarray*} \end{eqnarray}  %%%%%%%%%%%%%%%%
%%%%%%%%%%%%%%%%%%%%%%%%%%%%%%%%%%%%%%%%%%%%%%%%%%%%%%%%%%%%%%%%%%%%
\def\lan{\langle}
\def\ran{\rangle}
\def\p{\pi}
\def\ifmath#1{\relax\ifmmode #1\else $#1$\fi}%
\def\rc{\ifmath{{\mathrm{c}}}}
\def\cut{\ifmath{{\mathrm{cut}}}}
\def\rF{\ifmath{{\mathrm{F}}}}
\def\rK{\ifmath{{\mathrm{K}}}}
\def\rp{\ifmath{{\mathrm{p}}}}
\def\rt{\ifmath{{\mathrm{t}}}}
\def\LAB{\ifmath{{\mathrm{LAB}}}}
\def\cut{\ifmath{{\mathrm{cut}}}}
\def\beq{\begin{equation}}
\def\eeq{\end{equation}}

\newcommand{\chisq}[1]{$\chi^{2}_{#1}$}
\newcommand{\etap}{\eta^{\prime}}
\newcommand{\pip}{\pi^{+}}
\newcommand{\pim}{\pi^{-}}
\newcommand{\gam}{\gamma}
\newcommand{\piz}{\pi^{0}}
\newcommand{\rhoz}{\rho^{0}}
\newcommand{\az}{a_{0}(980)}
\newcommand{\fz}{f_{0}(980)}
\newcommand{\pipm}{\pi^{\pm}}
\newcommand{\psip}{\psi^{\prime}}
\newcommand{\psipp}{\psi^{\prime\prime}}
\newcommand{\jpsi}{J/\psi}
\newcommand{\ar}{\rightarrow}
\newcommand{\GeV}{GeV/$c^2$}
\newcommand{\MeV}{MeV/$c^2$}
\newcommand{\br}[1]{\mathcal{B}(#1)}
\newcommand{\cinst}[2]{$^{\mathrm{#1}}$~#2\par}
\newcommand{\crefi}[1]{$^{\mathrm{#1}}$}
\newcommand{\crefii}[2]{$^{\mathrm{#1,#2}}$}
\newcommand{\crefiii}[3]{$^{\mathrm{#1,#2,#3}}$}
\newcommand{\HRule}{\rule{0.5\linewidth}{0.5mm}}

\title{\Large \boldmath \bf Evidence for $\psip$ decays into $\gamma \pi^0$ and $\gamma\eta$}

%\author{Author list}
%\begin{small}
\author{
%\begin{small}
%\begin{center}
{\small M.~Ablikim$^{1}$, M.~N.~Achasov$^{5}$, L.~An$^{9}$, Q.~An$^{35}$, Z.~H.~An$^{1}$, J.~Z.~Bai$^{1}$, R.~Baldini$^{17}$, Y.~Ban$^{22}$, J.~Becker$^{2}$, N.~Berger$^{1}$
% ...
\\
\vspace{0.2cm}
(BESIII Collaboration)\\
\vspace{0.2cm} {\it
$^{1}$ Institute of High Energy Physics, Beijing 100049, P. R. China\\
$^{2}$ Bochum Ruhr-University, 44780 Bochum, Germany\\
% ...
$^{5}$ G.I. Budker Institute of Nuclear Physics SB RAS (BINP), Novosibirsk 630090, Russia\\
% ...
$^{9}$ Guangxi University, Naning 530004, P. R. China\\
% ...
$^{17}$ Laboratori Nazionali di Frascati - INFN, 00044 Frascati, Italy\\
% ...
$^{22}$ Peking University, Beijing 100871, P. R. China\\
% ...
$^{35}$ University of Science and Technology of China, Hefei 230026, P. R. China\\
% ...
}}
%\end{center}
%\end{small}
\vspace{0.4cm} }
%\end{small}

%\collaboration{${\it BESIII}$  Collaboration}
%\date{\today}

\collaboration{BES Collaboration}

\begin{abstract}
The decays $\psip \rightarrow \gamma \pi^0$, $\gamma \eta$ and
$\gamma\eta^\prime$ are studied using data collected with the BESIII
detector at the BEPCII $e^+e^-$ collider. Processes
$\psip\to\gamma\pi^0$ and $\psip\to\gamma\eta$ are observed for the
first time with signal significances of 4.6$\sigma$ and 4.3$\sigma$,
respectively. The branching fractions are determined to be:
$\mathcal{B}(\psip \to\gamma\pi^0) =
(1.58\pm0.40\pm0.13)\times10^{-6}$, $\mathcal{B}(\psip\to\gamma\eta)
= (1.38\pm0.48\pm0.09)\times10^{-6}$, and $\mathcal{B}(\psip
\to\gamma\eta^\prime) = (126\pm3\pm8)\times10^{-6}$, where the first
errors are statistical and the second ones systematic.
\end{abstract}


\maketitle

%%%%%%%%%%%%%%%%%%%%%%%%%%%%%%%%%%INTRODUCTION%%%%%%%%%%%%%%%%%%%%%%%%%%%%%%%%%%%%%%%%%%%%%%%%%%%%%%
The study of vector charmonium radiative decay to a neutral
pseudoscalar meson $P=(\pi^0, \eta, \eta^{\prime})$ provides important
tests for various phenomenological mechanisms, such as the vector meson
dominance model (VDM)~\cite{cz-report,korner-1983,intman_vdm},
two-gluon couplings to $q\bar{q}$ states~\cite{korner-1983}, mixing of
$\eta_c-\eta^{(\prime)}$~\cite{fritzsch-1977,chao-1990}, and
final-state radiation by light quarks~\cite{cz-report}. Direct
contributions from the continuum through a virtual photon: $e^+e^-
\rightarrow \gamma^* \to \gamma P$ are relevant to the decays of
$J/\psi$, $\psip$ and $\psipp$ to $\gamma P$ as discussed recently in
Ref.~\cite{rosner_prd_2010}. Furthermore, the possible interference
between the charmonium decays and continuum process may play a key
role in understanding the difference between $J/\psi$ and $\psip$
decays into $\gamma P$~\cite{yuancz-wangp-mo}.

For $P=\eta$ and $\eta^\prime$, the ratio $R_{J/\psi}\equiv
\mathcal{B}(J/\psi \to \gamma \eta)/\mathcal{B}(J/\psi \to \gamma
\eta^\prime)$ can be predicted by first order perturbation
theory~\cite{cz-report}. The analogous ratio ($R_{\psip}$) can be defined
for $\psip$ radiative decays into $\eta$ and $\eta^\prime$, and
$R_{\psip} \approx R_{J/\psi}$ is
expected~\cite{cleo-c-2009-gp}. Recently, the CLEO Collaboration
reported measurements for the decays of $J/\psi$, $\psip$ and $\psipp$
to $\gamma P$~\cite{cleo-c-2009-gp}, and no evidence for $\psip \to
\gamma \eta$ or $\gamma \pi^0$ was found. Therefore they obtain
$R_{\psip} \ll R_{J/\psi}$ with $R_{\psip}<1.8\%$ at the 90\% C.L. and
$R_{J/\psi} = (21.1\pm0.9)\%$~\cite{cleo-c-2009-gp}. Such a small
$R_{\psip}$ is unanticipated, and it poses a significant challenge to
our understanding of the $c\bar{c}$ bound states.

The decay $\psip \to \gamma \pi^0$ is suppressed in QED because the photon can only be produced
from final state radiation off one of the quarks. It has also been described via the strong process
$\psip \rightarrow ggg\to \rho^*\pi^0$, $\rho^*\to \gamma$ in the VDM~\cite{intman_vdm}. In
Ref.~\cite{rosner_prd_2010}, the contribution from $\psip \to \gamma^* \to \gamma \pi^0$ is
calculated, and $\mathcal{B}(\psip \rightarrow \gamma \pi^0)\approx 2.19\times 10^{-7}$ is
obtained, which is compatible to the VDM contribution and does not contradict the upper limit of
$5.0\times 10^{-6}$ (at the 90\% C.L.) reported by the CLEO Collaboration~\cite{cleo-c-2009-gp}.
The $\gamma^* -\gamma -\pi^0$ vertex was shown~\cite{brodsky-1980-1981} to be characterized by a
form factor $F(Q^2)$, where $Q^2\equiv -q^2$ and $q$ is the four-momentum of the virtual photon
$\gamma^*$. By using $e^+e^- \to e^+e^- \pi^0$, the form factor was measured in the
CLEO~\cite{cleo-ggpi} and {\it BABAR}~\cite{babar-ggpi} experiments for spacelike nonasymptonic
momentum transfer in the range $|q^2|=1.6-8.0$ GeV$^2$ and $4-40$ GeV$^2$, respectively. The
$e^+e^- \to \psip/\gamma^* \to \gamma \pi^0$ process will be very useful in testing the form factor
for timelike photons $Q^2=-q^2<0$~\cite{rosner_prd_2010}.

%%%%%%%%%%%%%%%%%%%%%%%%%%%%%%%%%%%%%%%%%%%%%%%%%%%%%%%%%%%%%%%%%%%%%%%%%%%%%%%%%%%%%%%%%%%%%%%%%%%%
In this Letter, $\psip \to \gamma \pi^0$ is studied using $\pi^0 \to
\gamma \gamma$ decay, $\psip \to \gamma \eta$ is measured using $\eta
\to \pi^+\pi^-\pi^0$ and $\eta \to \pi^0\pi^0\pi^0$ with $\pi^0 \to
\gamma \gamma$, and $\psip \to \gamma \etap $ is studied using
$\etap\to \gamma \pi^+\pi^-$ and $\etap\to \pi^+\pi^-\eta$ with $\eta
\to \gamma\gamma$. The analyses use a data sample of 156.4
$\rm{pb}^{-1}$ collected at the $\psip$ peak with the BESIII detector
operating at BEPCII~\cite{ref:bes3nim,ref:bes3phy}. By measuring
the production of multihadronic events, the number of $\psip$
decays is found to be $(1.06\pm0.04)\times10^8$ \cite{ref:psiptotnumber}. An
independent data sample of 42.6 $\rm{pb}^{-1}$ taken at
$\sqrt{s}=3.65$ GeV is utilized to determine the potential background
contribution from the continuum.

BEPCII is a double-ring $e^+e^-$ collider designed to provide
$e^+e^-$ beams with a peak luminosity of $10^{33}
~\rm{cm}^{-2}\rm{s}^{-1}$ at a beam current of 0.93 A. The cylindrical
core of the BESIII detector consists of a helium-based main drift
chamber (MDC), a plastic scintillator time-of-flight system (TOF), and
a CsI(Tl) electromagnetic calorimeter (EMC), which are all enclosed in
a superconducting solenoidal magnet providing a 1.0 T magnetic
field. The solenoid is supported by an octagonal flux-return yoke with
resistive plate counter muon identifier modules interleaved with
steel. The acceptance of charged particles and photons is 93\% over
4$\pi$ stereo angle, and the charged-particle momentum and photon
energy resolutions at 1 GeV are 0.5\% and 2.5\%, respectively.

The BESIII detector is modeled with a Monte Carlo (MC) simulation
based on \textsc{geant}{\footnotesize
4}~\cite{ref:geant4,ref:geant4_2}. \textsc{
evtgen}~\cite{ref:bes3gen} is used to generate $\psip\ar\gamma\piz,
\gam\eta$, $\gam\etap$ events, where the angular distribution of the
radiative photon from $\psip$ decay is $1+\rm{cos^2 \theta}$ in the
$\psip$ frame. The decay $\eta\ar\pip\pim\piz$ is generated
according to the Dalitz distribution measured
in~\cite{ref:etaDalitz} and $\etap\ar\gamma\pip\pim$ is simulated
assuming it is mediated by $\rho^0\ar\pip\pim$, while the decays of
$\eta \to \pi^0\pi^0 \pi^0$ and $\etap \to \pi^+\pi^- \eta$ are
generated with phase space. $\psip$ decays are simulated by the MC
event generator \textsc{kkmc} \cite{ref:kkmc} with known decays
modeled by the \textsc{evtgen} according to the branching fractions
provided by the Particle Data Group (PDG)~\cite{PDG}, and the
remaining unknown decay modes generated with \textsc{
lundcharm}~\cite{ref:bes3gen}.

Charged tracks in BESIII are reconstructed using MDC hits. To
optimize the momentum measurement, we select tracks in the polar
angle range $|\cos \theta| <0.93$ and require that they pass within
$\pm10$ cm from the Interaction Point (IP) in the beam direction and
within $\pm1$ cm of the beam line in the plane perpendicular to the
beam. All the charged tracks are assumed to be pions, and particle
identification (PID) is not required, except in $\etap\ar\gamma\rho$
where the $dE/dx$ information has been used to suppress QED
background, most of which is from $e^+e^- \to e^+e^- \gamma$. Either
zero or two tracks with net charge zero are required for the final
$\pi^0/\eta/\etap$ decay products.

Electromagnetic showers are reconstructed by clustering EMC
crystal energies. The energy deposited in nearby TOF counters is
included to improve the reconstruction efficiency and energy
resolution. Showers identified as photon candidates must satisfy
fiducial and shower-quality requirements. For the $\psip \to \gamma
\eta~\rm{and}~\gam\etap$ analyses, the photon candidate showers are
reconstructed from both the barrel and endcap of the EMC, and showers
from barrel region $(|\cos \theta|<0.8)$ must have a minimum energy
of 25 MeV, while those in the endcaps $(0.86<|\cos \theta|<0.92)$
must have at least 50 MeV. The showers in the angular range between
the barrel and endcap are poorly reconstructed and excluded from the
analyses. To exclude showers from charged particles, a photon must be
separated by at least $10^\circ$ from any charged track.  The EMC
cluster timing requirements are used to suppress electronic noise
and energy deposits unrelated to the event.

Events with the decay modes shown in Table \ref{sum:eff_yield} are
selected. Every particle in the final state must be explicitly found,
and their vertex must be consistent with the measured beam spot. The
sum of four-momenta of all particles is constrained to the known
$\psip$ mass~\cite{PDG} and initial $e^+e^-$ three-momentum in the lab
frame. The vertex and full event four-momentum kinematic fits must
satisfy $\chi^2_{\rm{Vx}} <100$ and $\chi^2_{\rm{4C}} <40$ ,
respectively. For $\eta\to 3 \pi^0$, a looser restriction of
$\chi^2_{\rm{4C}} <90$ is applied to increase efficiency. Further
selections are based on four-momenta from the kinematic fit. In
$\eta/\etap$ channels, photon pairs are used to reconstruct $\piz$ or
$\eta$ candidates if their invariant mass satisfies $M_{\gam\gam} \in
$ (120, 150) \MeV~ or (515, 565) \MeV, respectively.
%  \vspace{-0.0cm}

\begin{table}[hbtp]
  \footnotesize
  \caption{For each decay mode, the number of signal events ($N_S$),
  the number of scaled continuum background events ($N_C$) in the
  signal region, the number of expected background events from $\psip$
  decays ($N_{R}$) in the signal region, and the MC efficiency
  ($\varepsilon$) for signal are given. The error on $N_S$ is only the
  statistical error, and the signal region is defined to be within
  $\pm3\sigma$ from the nominal $\pi^0$, $\eta$, and $\etap$
  masses.}
  \label{sum:eff_yield}
  \begin{center}
     \renewcommand{\arraystretch}{1.1}
     \begin{tabular}{l|cccc}
        \hline\hline
        Modes    ($\psip\ar\gam X$)                     & $N_S$    ~~ & $N_C$ & ~~$N_R$  &  ~~$\epsilon$(\%)\\
        \hline

        $  \psip\ar\gam\piz$                            & $37.4\pm9.5$ & 63.5 & 1.8 & 21.4  \\

        $  \psip\ar\gam\eta(\pip\pim\piz)$              & $8.9\pm3.6$  & 2.2  & 0.0 & 21.0  \\

        $  \psip\ar\gam\eta(\piz\piz\piz)$              & $3.8\pm2.3$  & 0.0  & 1.2 & 10.7  \\

        $  \psip\ar\gam\eta'[\pip\pim\eta(\gam\gam)]$   & $586\pm25$   & 0.0  & 4.7 & 27.1  \\

        $  \psip\ar\gam\eta'(\pip\pim\gam)$             & $1640\pm44$ & 179.3 & 111.7 & 41.0  \\

        \hline\hline
      \end{tabular}
      \vspace{-0.6cm}
  \end{center}
\end{table}

For the $\psip \to \gamma \pi^0$ analysis, the primary background
comes from the continuum process $e^+e^-\ar \gamma \gamma (\gamma)$,
where the two energetic photons are distributed in the forward and
backward regions. We require that photon candidate showers lie in the
barrel region of the EMC to suppress this background. Since $\pi^0$
mesons decay isotropically, the angular distribution of photons from
$\piz$ decays is flat in the $\piz$ helicity frame. However, continuum
background events accumulate near $\cos \theta_{\rm{decay}}=\pm 1$,
where $\theta_{\rm{decay}}$ is the angle of the decay photon in the
$\piz$ helicity frame~\cite{ref:psiptotnumber}. To further suppress
continuum background, we require $|\cos
\theta_{\rm{decay}}|<0.5$. Another potentially serious background
comes from $e^+e^-\to \gamma \gamma$, in which one $\gamma$ converts
into an $e^+e^-$ pair in the outer part of the MDC. If the track
finding algorithm fails to find the track, the two showers in the EMC
are identified as isolated photons without associated charged
tracks. To suppress this background, the number of MDC hits,
$N_{hits}$, is counted in the sector between the two radial lines
connecting the IP and the two-shower positions in the EMC.  To take
the EMC spacial resolution into account, the sector is extended by 3.5
degrees on both sides. Because of high beam related background level,
the hits in the inner 8 MDC layers are not counted in
$N_{hits}$. Figure~\ref{fig:mdchits} (a) shows the $N_{hits}$
distribution. As shown in Fig.~\ref{fig:mdchits} (b), background from
continuum event $\gamma$ conversions $e^+e^-\ar \gamma \gamma$
accumulates in the low mass region. After requiring $N_{hits}\leq 10$,
this background is reduced dramatically, while there is still an
accumulation of events at the $\pi^0$ mass, as shown in
Fig.~\ref{fig:mdchits} (c).

%\vspace{-0.2cm}
\begin{figure}[hbtp]
   \footnotesize
    \includegraphics[width=0.5\textwidth]{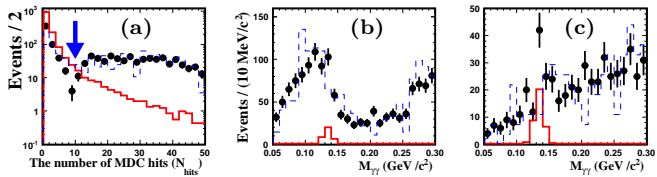}
    \put(-213, 58){  \bf (a)}
    \put(-126, 58){  \bf (b)}
    \put(-45,  58){  \bf (c)}
    \vspace{-0.4cm}
    \caption{The distribution of (a) the number of MDC hits, and the
      two-photon invariant mass distributions for $\gamma \pi^0$ final
      states (b) without and (c) with the $N_{hits}\leq 10$
      requirement. Solid histograms are MC simulated signal for $\psip
      \to \gamma \pi^0$; dashed histograms are the luminosity-scaled
      continuum data, and points are $\psip$ data. The solid arrow
      indicates the requirement on $N_{hits}$. \label{fig:mdchits}}
\end{figure}
%\vspace{-0.2cm}

\begin{figure}[hbtp]
%\vspace{-1.0cm}
\begin{minipage}[t]{0.45\linewidth}
    \centering
    \includegraphics[width=0.94\linewidth, height=0.7\textwidth]{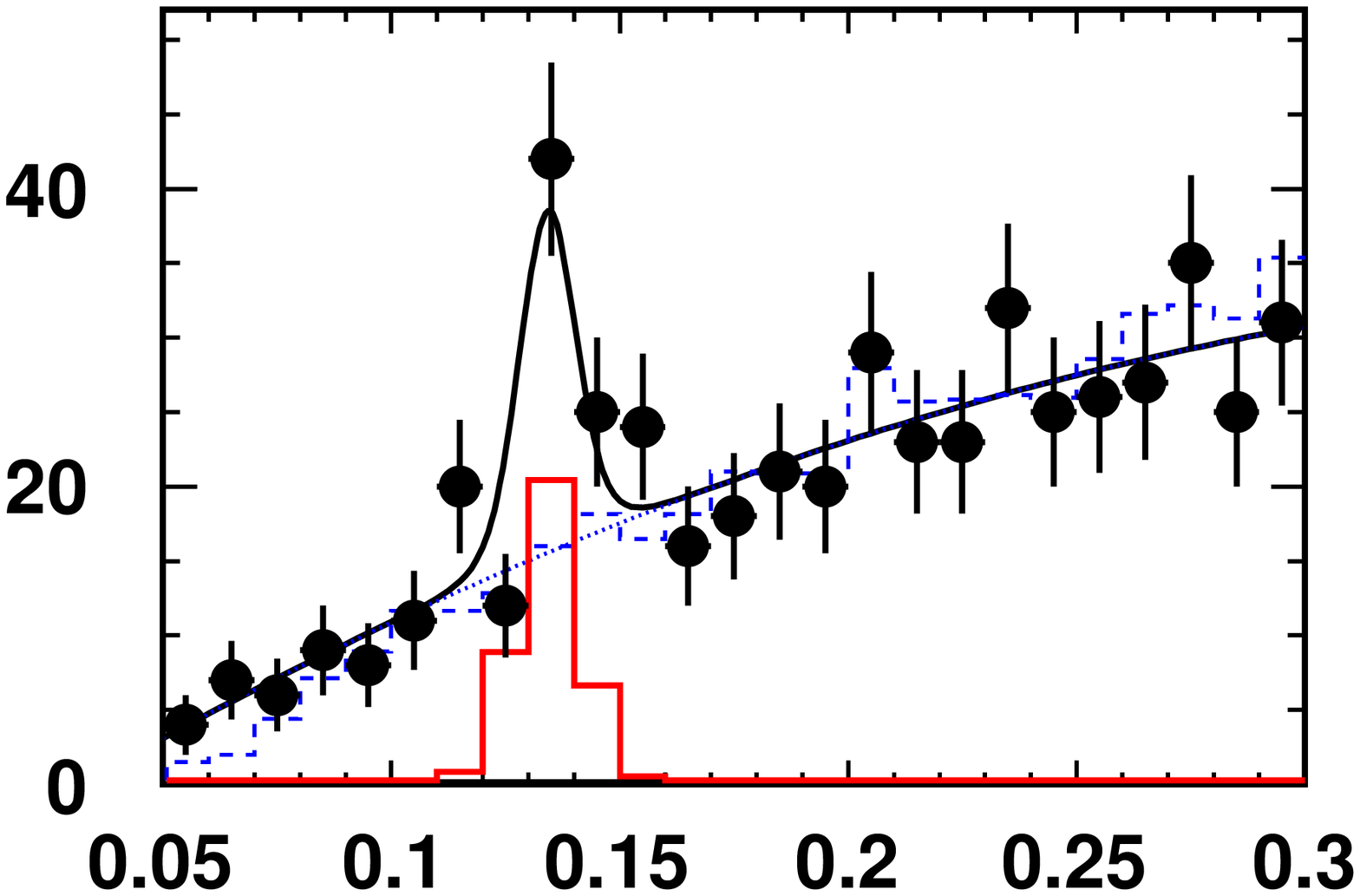}
\end{minipage}
\put(-90, 55){ \bf (a)}
%\hspace{-0.6cm}
\begin{minipage}[t]{0.45\linewidth}
    \centering
    \ \
\end{minipage}

\vspace{-0.8cm}
\begin{minipage}[t]{0.45\linewidth}
   \centering
    \includegraphics[width=0.95\linewidth, height=1.4\textwidth]{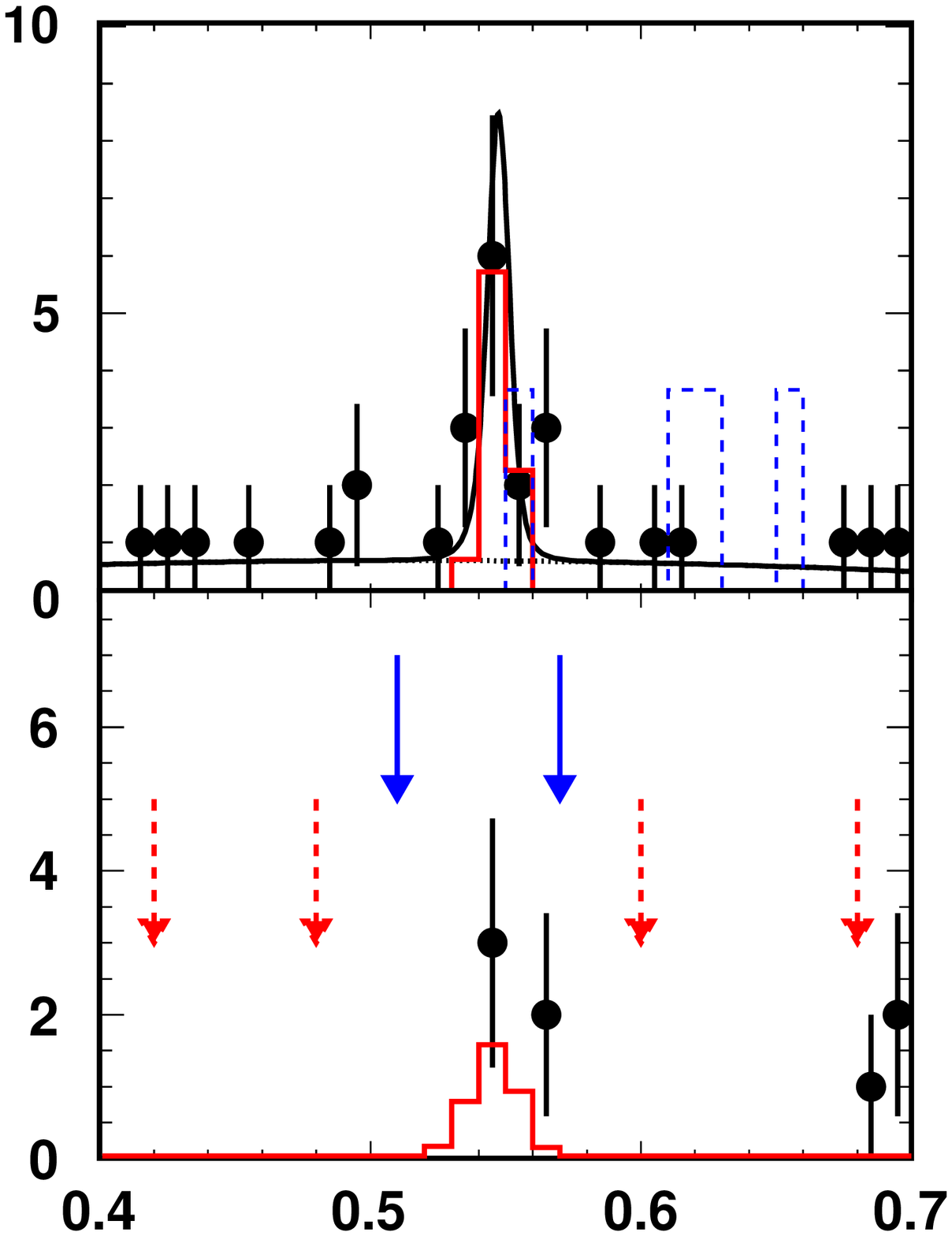}
\end{minipage}
\hspace{-0.4cm}
\begin{minipage}[t]{0.45\linewidth}
    \centering
    \includegraphics[width=0.95\linewidth, height=1.4\textwidth]{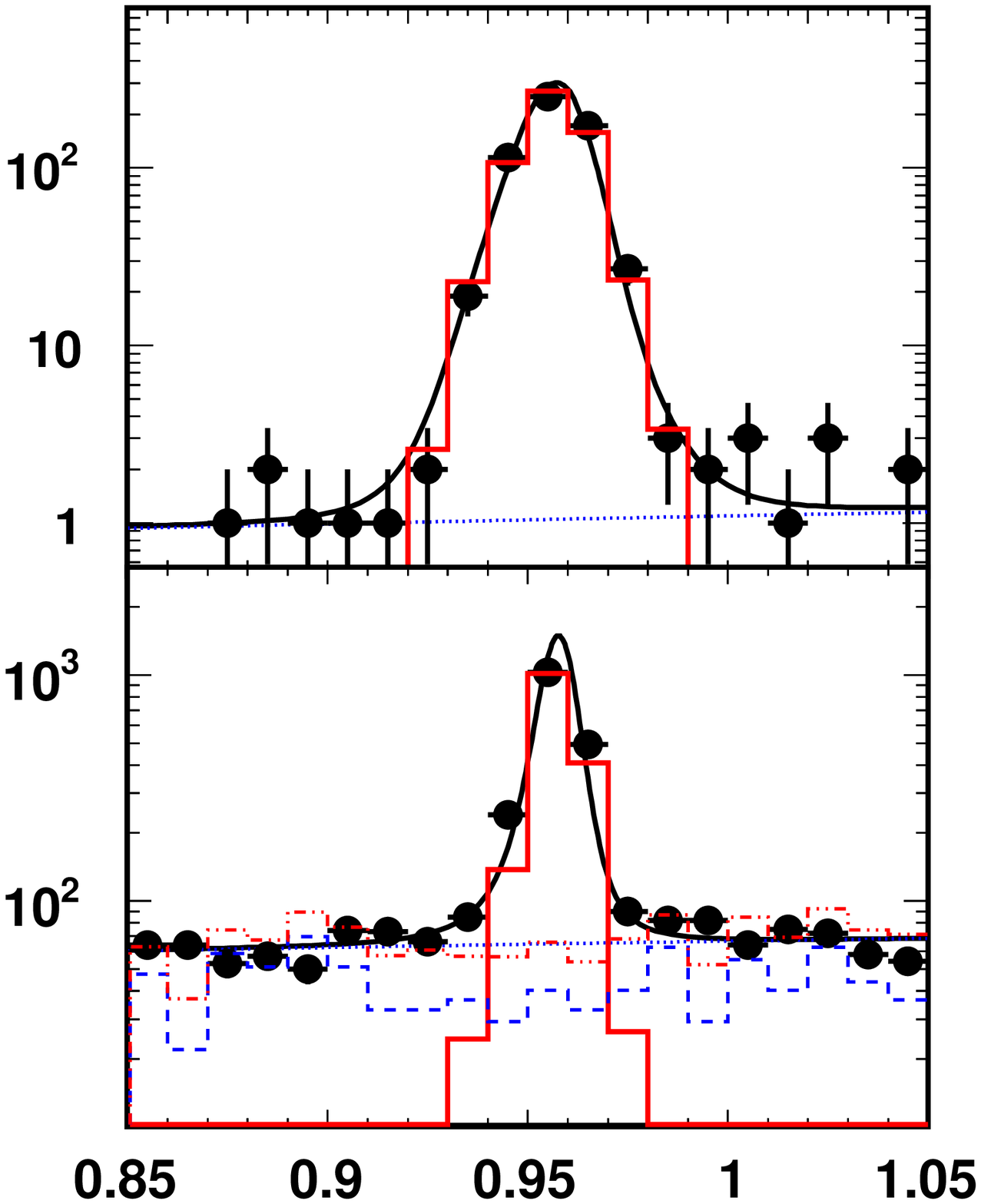}
    \put(-195,138){ \bf (b)}
    \put(-195,81){ \bf (c)}
    \put(-85, 138){ \bf (d)}
    \put(-85, 81){ \bf (e)}
    \put(-150,15){ \textbf{Mass (\GeV)}}
    \put(-225,80){\rotatebox{90}{ \textbf{Events/(10 \MeV)}}}
    \end{minipage}
\vspace{-0.5cm} \caption{Mass distributions of the pseudoscalar meson candidates
  in $\psip\ar\gam P$: (a) $\gam\piz$, (b) $\gam\eta(\pip\pim\piz)$;
  (c) $\gam\eta(3\piz)$; (d) $\gam\etap[\pip\pim\eta(\gam\gam)]$; and
  (e) $\gam\etap(\gam\pip\pim)$. The crosses are data, the solid
  histograms are MC simulated signal, and the dashed lines are the
  continuum backgrounds. Fits are shown as solid lines, background
  polynomials as dotted lines. In (c), the arrows indicate nominal
  selection criteria, while the dashed arrows show sidebands. In (e),
  the dot-dot-dash line shows the sum of the continuum background and
  the expected background from $\psip$ decays.}
\label{fit:all_data} \vspace{-0.5cm}
\end{figure}

After applying the above selection criteria, the mass spectra of
$\piz,\eta,~\rm{and}~\etap $ candidates are shown in
Fig.~\ref{fit:all_data}. An unbinned maximum likelihood (ML) fit is
used for each analysis to determine the event yields except for
$\psip\to \gamma \eta(\pi^0\pi^0\pi^0)$. The signal probability
density function (PDF) in each mode is obtained from MC simulation.
The shape for the continuum background is described by a second
order Chebychev polynomial function, and the yield and its PDF
parameters are floated in the fit. The fitting ranges for
$\pi^0,\eta ~\rm{and}~\etap$ are 0.05--0.30 \GeV, 0.40--0.70 \GeV,
and 0.85--1.05 \GeV, respectively. The signal yield for
$\psip\ar\gamma\eta(\piz\piz\piz)$ is determined directly by
counting the number of events in the signal region, which is in
0.51--0.57 GeV/$c^2$, about 3 standard deviations from the nominal
value of the $\eta$ mass~\cite{PDG} as shown in
Fig.~\ref{fit:all_data} (c), while the sideband regions are defined
as 0.42--0.48 GeV/$c^2$ and 0.60--0.66 GeV/$c^2$. The signal yields
and the efficiencies are summarized in Table~\ref{sum:eff_yield}.

The backgrounds remaining after event selection can be divided into
two categories. One is from $\psip$ decays, which can be studied
using a sample of $10^8$ MC-simulated inclusive $\psip$ events.  The
other is from non-resonant processes or initial state radiation to
low mass resonances, which can be studied using the continuum data
sample collected at a center of mass energy of 3.65 GeV.  The
expected backgrounds from $\psip$ decays are listed in Table
\ref{sum:eff_yield}, where the number of background events is the
number in the signal region, which is defined as within $\pm3
\sigma$ from the nominal $\pi^0$, $\eta$ and $\etap$ masses. For
$\psip\ar\gamma\piz$, the normalized number of events from
$\psip\ar\gamma\piz\piz$ is 1.8 in the $\pi^0$ signal range. For
$\psip\ar\gam\eta(\piz\piz\piz)$, there are 1.2 events from the
decay of $\psip\ar\gamma\eta(\gamma\gamma)\eta(3\piz)$. For
$\psip\ar\gam\etap(\gam\pip\pim)$, the main background from $\psip$
decays is $\psip\ar\rho^0\pi^0$ which contributes a smooth
background. The QED backgrounds for
$\psip\ar\gam\eta'(\pip\pim\gam)$ are from
$e^+e^-\ar\gamma\mu^+\mu^-$ and $e^+e^-\ar\gamma e^+e^-$, and both
of them give a smooth background under the $\etap$ signal peak.  For
$\psip\ar\gamma\piz$, a smooth background is contributed from
$e^+e^-\ar\gam\gam(\gam)$ events.  The cross section for $e^+e^-
\rightarrow \gamma^* \rightarrow \gamma \piz(\gam\eta)$ has been
estimated using data collected at $\sqrt{s}=3.65$ GeV, and the upper
limit on the cross section is less than 0.14(0.68) pb at the 90\%
C.L. Since the continuum cross section is small, we neglect possible
interference between $\psip\ar \gamma \piz(\eta)$ signal and
continuum $\gamma\piz(\eta)$. All the backgrounds are summarized in
Table \ref{sum:eff_yield}.

\begin{table*}[hbtp]
%  \footnotesize
\begin{center}
\vspace{-0.0cm} \caption{ {\boldmath{Summary of systematic errors (\%).}} } \label{tab:sys_all}
  \vspace{0.3cm}
    \renewcommand{\arraystretch}{1.1}
     \begin{tabular}{l|ccccc}
        \hline\hline

        Sources  &~~$\pi^0$ & ~~$\eta\to\pi^+\pi^-\pi^0$ & ~~$\eta\to 3 \pi^0$
        & ~~$\etap\to\pi^+\pi^-\eta(\gamma\gamma)$ & ~~$\etap\to\pi^+\pi^-\gamma$\\ \hline

        MDC track finding               &  --   &  4   &   --  & 4   & 4   \\

        %Radiative $\gamma$ Detection     & 0.25  &  0.25  &  0.25 & 0.25  & 0.25  \\

        Photon detection                &  2  &  2   &  6  & 2   & 1   \\

        $\pi^0(\eta)$ reconstruction    &  --   &  1   &  3  & 1   & --    \\

        4C kinematic fit                &  1  &  3   &  0  & 3   & 2   \\

        Background shape                &  4.8  &  6.4   &  --   & 0   & 1   \\

        Number of $\psip$               &  4  &  4   &  4  & 4   & 4   \\

        Cited branching fractions       &  0  &  1.2   &  0.7  & 1.7   & 1.7   \\

        MDC hits                        &  3  &   --   &  --   &  --   &  --   \\

        Number of photons               &  4  &   --   &  --   &  --   &  --   \\

        Total                           &  8.3  &  9.4   &  7.8  & 7.0   & 6.4   \\

        \hline\hline
      \end{tabular}
      \vspace{-0.5cm}
\end{center}
\end{table*}

The systematic uncertainties for these measurements are summarized in
Table~\ref{tab:sys_all}. The uncertainties due to MDC track finding
and photon detection are 2\% per charged track and 1\% per low energy
photon. The uncertainty of detecting the high energy photon is less
than 0.25\% which can be neglected. The systematic
errors from $\piz$ ($\eta$) reconstruction is determined to be 1\% per
$\piz$ ($\eta$) by using a high purity control sample of
$J/\psi\ar\piz\bar{p}p$ ($J/\psi\ar\eta\bar{p}p$) decay. The
uncertainties due to kinematic fits have been estimated using the
control samples with the same event topologies as those in the signal
cases, i.e. the same number of charged tracks and same number of
photons. The systematic uncertainties due to the $dE/dx$ requirements to
identify charged pions in the $\psip\ar \gamma \etap(\gam\pip\pim)$
 and $N_{\gam}$ in the $\psip\ar\gam\piz$ are studied by
using the control samples of $\jpsi\ar\rho\pi$ and
$\jpsi\ar\gam\eta(\gam\gam)$, respectively, with and without applying
these requirements.
% Then, the variations in the signal yields are
%taken as the systematic errors.

In $\psip\ar \gamma \pi^0$, the uncertainty due to the requirement
on the MDC hits, $N_{hits}\leq 10$, is studied using a sample of
$\jpsi\ar\gam\eta, \eta\ar\gam\gam$ events. The ratios of events
with and without the requirement on the number of MDC hits are
obtained for both data and MC simulation. Taking the difference of
opening angle between $J/\psi\ar\gamma\eta(\gamma\gamma)$ and
$\psip\ar\gamma\piz$ into account, the difference 3\% is considered
as the systematic error for the measurement of $\psip\ar \gamma
\pi^0$ and is due to the difference in the noise in the MDC for data
and MC simulation.

The uncertainty due to the background shape has been estimated by
varying the PDF shape and fitting range in the ML fit. For the
intermediate decays, the $\eta(\etap)$ branching fractions and
uncertainties from the PDG fit~\cite{PDG} are used. The total
relative systematic errors on these measurements are 8.3\%, 9.4\%,
7.8\% , 7.0\%, and 6.4\% for $\psip\ar\gamma \pi^0$, $\psip\ar\gamma
\eta (\pi^+\pi^-\piz)$, $\psip\ar\gamma\eta(\pi^0\pi^0\pi^0)$,
$\psip\ar \gamma \etap[\pi^+\pi^-\eta(\gamma\gamma)]$, and
$\psip\ar\gamma \etap(\pi^+\pi^-\gamma)$, respectively, as
summarized in Table~\ref{tab:sys_all}.

The branching fractions of $\psip$ decays to $\gam$ and a
pseudoscalar meson are listed in Table~\ref{final_br}. Taking the
common systematic errors into account, the combined measurements for
$\psip\ar\gamma\eta, ~\gamma\etap$ modes are obtained. The
PDG~\cite{PDG} values are also shown in Table~\ref{final_br}. With
considering the background shape uncertainty, we find the signal
significance for $\psip\ar\gamma\piz(\gamma\eta)$ to be
4.6(4.3)$\sigma$, as determined by the ratio of the maximum
likelihood value and the likelihood value for a fit where the signal
contribution is set to zero.

%\vspace{-0.3cm}
\newcommand{\rb}[1]{\raisebox{2.0ex}[0pt]{#1}}
\begin{table}[hbtp]
  \footnotesize
  \caption{Branching fractions ($10^{-6}$) from this analysis, where the first errors are statistical
  and the second ones are systematic, and the comparison with the PDG values~\cite{PDG}.}
  \vspace{-0.4cm}
  \label{final_br}
  \begin{center}
     \renewcommand{\arraystretch}{1.1}
     \begin{tabular}{l|ccc}
        \hline\hline
        Mode         &   BESIII  & Combined BESIII  & PDG  \\
        \hline

        $\psip\ar\gam\piz$   & $1.58\pm0.40\pm0.13$ & $1.58\pm0.40\pm0.13$ & $\leq5$ \\

        $\psip\ar\gam\eta(\pip\pim\piz)$  & $1.78\pm0.72\pm0.17$ &                      &         \\

        ~~$\ar\gam\eta(\piz\piz\piz)$     & $1.07\pm0.65\pm0.08$ & \multicolumn{1}{c}{\rb{$1.38\pm0.48\pm0.09$}} & \multicolumn{1}{c}{\rb{$\leq2$}}\\

        $\psip\ar\gam\etap(\pip\pim\eta)$ & $120\pm5\pm8$        &                      &         \\

        ~~$\ar\gam\etap(\pip\pim\gam)$    & $129\pm3\pm8$        & \multicolumn{1}{c}{\rb{$126\pm3\pm8$}} & \multicolumn{1}{c}{\rb{$121\pm8$}}\\

        \hline\hline

      \end{tabular}
      \vspace{-0.5cm}
  \end{center}
\end{table}

In summary, we have measured branching fractions for
$\psip\ar\gamma\piz$, $\psip\ar \gamma \eta$ and $\psip\ar \gamma
\etap$ decays. For the first time, we find evidence for the
$\psip\ar\gamma\piz$ and $\psip\ar \gamma \eta$ decays with signal
significances of 4.6$\sigma$ and 4.3$\sigma$, respectively. The
evidence for $\psip\ar \gamma \piz$ will yield an experimental
constraint on the $\gamma^*\ar \gamma \piz$ vertex in the timelike
regime at $|q^2|=m^2_{\psip}$~\cite{rosner_prd_2010}. For the ratio
of $\eta~\rm{and}~\etap$ production rates from $\psip$ decays, we
obtain $R_{\psip}=(1.10\pm0.38\pm0.07)\%$, where the statistical and
systematic uncertainties from the input branching fractions as
listed in Table~\ref{final_br} have been combined in quadrature
after accounting for common systematic errors. This ratio is the
first measurement, and it is below the 90\% C.L. upper bound
determined by the CLEO Collaboration~\cite{cleo-c-2009-gp}. The
corresponding $\eta - \etap$ production ratio for the $J/\psi$
resonance was measured to be
$R_{J/\psi}=(21.1\pm0.9)\%$~\cite{cleo-c-2009-gp}. $R_{\psip}$ is
smaller than $R_{J/\psi}$ by an order of magnitude.
% Such a small value of $R_{\psip}$ poses a great
%challenge to our understanding of the decay properties of the
%charmonium states.

The BESIII collaboration thanks the staff of BEPCII and the
computing center for their hard efforts. This work is supported in
part by the Ministry of Science and Technology of China under
Contract No. 2009CB825200; National Natural Science Foundation of
China (NSFC) under Contracts Nos. 10625524, 10821063, 10825524,
10835001, 10935007, 10979008; the Chinese Academy of Sciences (CAS)
Large-Scale Scientific Facility Program; CAS under Contracts Nos.
KJCX2-YW-N29, KJCX2-YW-N45; 100 Talents Program of CAS; Istituto
Nazionale di Fisica Nucleare, Italy; Russian Foundation for Basic
Research under Contracts Nos. 08-02-92221, 08-02-92200-NSFC-a;
Siberian Branch of Russian Academy of Science, joint project No 32
with CAS; U. S. Department of Energy under Contracts Nos.
DE-FG02-04ER41291, DE-FG02-91ER40682, DE-FG02-94ER40823; University
of Groningen (RuG) and the Helmholtzzentrum fuer
Schwerionenforschung GmbH (GSI), Darmstadt; WCU Program of National
Research Foundation of Korea under Contract No.
R32-2008-000-10155-0.


\begin{thebibliography}{99}
%
%
\bibitem{cz-report} V.~L.~Chernyak and A.~R.~Zhitnitsky, Phys. Rep. {\bf 112}, 173 (1984).

\bibitem{korner-1983} J.~G.~K\"{o}rner {\em et al.}, Nucl. Phys. B {\bf 229}, 115 (1983).

\bibitem{intman_vdm} G. W. Intemann, Phys. Rev. D \textbf{27}, 2755 (1983).

\bibitem{fritzsch-1977} H.~Fritzsch and J.~D.~Jackson, Phys. Lett. B {\bf 66}, 365 (1977).

\bibitem{chao-1990} K.~T.~Chao, Nucl. Phys. B {\bf 335}, 101 (1990).

\bibitem{rosner_prd_2010} J.~L.~Rosner, Phys. Rev. D {\bf 79}, 097301 (2009).

\bibitem{yuancz-wangp-mo} P.~Wang {\em et al.},  Phys. Lett. B {\bf 593}, 89 (2004).

\bibitem{cleo-c-2009-gp} T. K. Pedlar {\em et al.} (CLEO Collaboration), Phys. Rev. D {\bf79}, 111101 (2009).

\bibitem{brodsky-1980-1981} G.~P.~Lepage and S.~J.~Brodsky, Phys. Rev. D {\bf 22}, 2157(1980);
S.~J.~Brodsky and G.~P.~Lepage, Phys. Rev. D {\bf 24}, 1808 (1981).

\bibitem{cleo-ggpi} J.~Gronberg {\it et al.} (CLEO Collaboration), Phys. Rev. D {\bf 57}, 33 (1998).

\bibitem{babar-ggpi} B.~Aubert {\it et al.} ({\it{BABAR}} Collaboration), Phys. Rev. D {\bf80}, 052002 (2009).

\bibitem{ref:bes3nim} M.~Ablikim {\it et al.} (BES Collaboration), Nucl. Instrum. Methods Phys. Res., Sect. A {\bf 614}, 345 (2010).

\bibitem{ref:bes3phy} ``Physics at BES-III", edited by K. T. Chao and Y. Wang [Int. J. Mod. Phys. A {\bf 24}, Suppl. 1, (2009)]

\bibitem{ref:psiptotnumber} M.~Ablikim {\it et al.} (BES Collaboration), Phys. Rev. D {\bf 81}, 052005 (2010).

\bibitem{ref:geant4} S. Agostinelli {\it et al.} (\textsc{geant}{\footnotesize
4} Collaboration), Nucl. Instrum. Methods Phys. Res. A {\bf 506}, 250 (2003).

\bibitem{ref:geant4_2} J. Allison {\it et al.}, IEEE Trans. Nucl. Sci. {\bf 53}, 270 (2006).

\bibitem{ref:bes3gen} R.~G.~Ping, Chinese Phys. C {\bf 32}, 599 (2008).

\bibitem{ref:etaDalitz} J.~G.~Layter {\it et al.}, Phys. Rev. D {\bf 7}, 2565 (1973).

\bibitem{ref:kkmc} S.~Jadach, B.~F.~L.~Ward, and Z.~Was, Comput. Phys. Commun. {\bf 130}, 260
(2000); Phys. Rev. D {\bf 63}, 113009 (2001).

%\bibitem{PDG}C. Amsler {\em et al.} (Particle Data Group), Phys. Lett. B {\bf667}, 1 (2008).

\bibitem{PDG}K. Nakamura {\it et al.}  (Particle Data Group), J. Phys. G {\bf 37}, 075021 (2010).

\end{thebibliography}
\end{document}